\documentclass[pra,twocolumn,showpacs,nofootinbib]{revtex4}

\usepackage{amsmath}    
\usepackage{graphicx}   
\usepackage{verbatim}   
\usepackage{color}      
\usepackage{subfigure}  
\usepackage{hyperref}   

\begin{document}

\title{Photoabsorption of attosecond XUV light pulses by two strongly laser-coupled autoionizing states}
\date{\today}
\author{Wei-Chun Chu and C.~D.~Lin}
\affiliation{J. R. Macdonald Laboratory, Department of Physics, Kansas State University, Manhattan, Kansas 66506, USA}
\pacs{32.80.Qk,32.80.Zb,32.80.Fb,42.50.Gy}

\begin{abstract}

We study theoretically the photoabsorption spectra of an attosecond XUV pulse by a laser-dressed atomic system. A weak XUV excites
an autoionizing state which is strongly coupled to another autoionizing state by a laser. The theory was applied to
explain two recent experiments [Loh, Greene, and Leone, Chem. Phys. \textbf{350}, 7 (2008); Wang, Chini, Chen, Zhang, Cheng, He,
Cheng, Wu, Thumm, and Chang, Phys. Rev. Lett. \textbf{105} 143002 (2010)]
where the absorption spectra of the XUV lights were measured against the time delay between the laser and the XUV. In another
example, we study an attosecond pulse exciting the $2s2p(^1P)$ resonance of helium which is resonantly coupled to the $2s^2(^1S)$
resonance by a moderately intense 540~nm laser. The relation between the photoabsorption spectra and the photoelectron spectra
and the modification of the transmitted lights in such an experiment are analyzed. The role of Rabi flopping between the two
autoionizing states within their lifetimes is investigated with respect to the laser intensity and detuning.

\end{abstract}

\maketitle

\long\def\symbolfootnote[#1]#2{\begingroup%
\def\thefootnote{\fnsymbol{footnote}}\footnote[#1]{#2}\endgroup}


\section{Introduction}\label{introduction}

Autoionization in an atomic system is understood as the decay of a bound state to one or more degenerate continua. It has long
been observed in photoabsorption, photoion, or photoelectron spectra where resonances are characterized by their detailed shapes as
given by Fano's configuration interaction theory~\cite{fano}. In such energy-domain measurements the lifetimes are deduced from the
energy widths of the resonances. With the development of ultrafast technologies in the past decade, in particular, the emergence
of single attosecond pulses (SAPs) in the extreme ultraviolet (XUV) region~\cite{krausz}, it has become possible to ask whether
one can use a very short pulse to populate an autoionizing state and watch its decay within the lifetime. This issue has been
addressed theoretically~\cite{wickenhauser, mercouris, zhao, chu10}. In Ref.~\cite{chu10}, we proposed a pump-probe scheme using
two attosecond XUV pulses where the time evolution of the energy profile of the autoionizing wave packet initiated by the pump is
followed by another attosecond pulse within the decay lifetime. However, such measurements are not yet possible because the
available attosecond pulses are still too weak. Instead, in actual experiments today, attosecond pulses are used only to excite an
autoionizing state which is then ``probed'' by an intense infrared (IR) field~\cite{drescher, gilbertson}. However, the XUV and
the IR pulses often overlap in time, and thus it is not appropriate to consider the IR only as a probe pulse, especially when the
IR is rather intense. Instead, it is preferable to consider such XUV+IR experiments as photoabsorption or photoionization of a
laser-dressed atom. If photoionization occurs in the energy region where the electron continuum spectrum is featureless, i.e.,
without bound states or resonances, then the presence of the IR can be incorporated using the approximate ``streaking''
model~\cite{kitzler}. A number of such experiments using SAPs and attosecond pulse trains (APTs) have been
performed~\cite{hentschel, remetter, goulielmakis, mauritsson} and explained based on the attosecond streaking model. However,
bound states and autoionizing states are prevalent in atomic and molecular systems. The presence of an intense laser is likely to
resonantly or near-resonantly couple XUV-initiated states with other bound or autoionizing states. In the limit of long pulses,
such strongly coupled systems have been widely investigated in the past few decades as a model three-level atomic system, and
interesting phenomena such as electromagnetically induced transparency (EIT)~\cite{harris, fleischhauer}, coherent population
trapping~\cite{arimondo} and others\cite{tamarat, fleischhauer} have been widely investigated. Moreover, they are generally
perceived as an all-optical control of several characteristics of the light beam propagating in the medium, such as the speed,
absorption, storage, and retrieval of light~\cite{hau, phillips, vudyasetu}. In most of these studies, the coupling laser and the
probe laser (or the pump laser) are limited to visible and infrared lights. However, the weak probe (or pump) laser can be extended
to the high-energy spectral region from XUV to X-rays. These light sources excite
inner-shell electrons or multiply excited states of an atom while the laser strongly couple these states to other nearby atomic
states. In fact, EIT-like phenomena have been reported in such systems~\cite{loh} and EIT-like theory has also been extended to
the systems where the two strongly coupled states are autoionizing states~\cite{lambropoulos, bachau, madsen, themelis}.
For this case with long pulses, one can measure both absorption and electron spectra by scanning the wavelength of the probe pulse.

With the emergence of SAPs, we recently generalize the three-level autoionizing systems for the situation
where both the XUV and the laser are short pulses~\cite{chu11}. Specifically, we modeled the experiment by Gilbertson \textit{et
al.}~\cite{gilbertson} where a 140-attosecond pulse with central energy of about 60~eV is used to excite the $2s2p(^1P)$ doubly
excited state of helium in the presence of a moderately strong few-femtosecond IR laser. The IR laser can couple
$2s2p$ to the $2p^2(^1S)$ doubly excited state. This appears to be a typical three-level system where one can investigate
the autoionization dynamics of $2s2p$ within its lifetime (17~fs), or more precisely, the use of IR to control autoionization
dynamics by changing the time delay between the two pulses. Unfortunately, due to the large IR ionization rate from the higher
$2p^2$ state, the strong IR coupling in this case mainly provides an efficient pathway to remove electrons from $2s2p$, thus
depleting the autoionized electrons which would be observed in the time delay spectra. In order to observe the change of
the resonance profile, spectral resolution much better than 0.1~eV is needed. In the measurement~\cite{gilbertson}, the electron
spectrometer has a resolution of 0.7~eV. Thus, the lifetime information has been retrieved, but no information on the
autoionization dynamics has emerged from this experiment.

To achieve better spectral resolution, measuring photoabsorption is more favorable. An XUV+IR experiment was reported by Wang
\textit{et al.}~\cite{wang} where the absorption spectra of the laser-coupled autoionizing states in argon were measured
against the time delay. In this paper, we extend our earlier theory~\cite{chu11} to obtain absorption spectra versus time delay to
explain the results in Ref.~\cite{wang}. Practically, the effect of the light propagation in the gas medium has to be
incorporated~\cite{gaarde}, which we will address in the future. The present theory is also applied to the transient absorption
spectra with pulse trains reported by Loh \textit{et al.}~\cite{loh}, where the lengths of the two light pulses, 30 and 42~fs,
were longer than the lifetimes of the autoionizing states of interest. To fully utilize the laser coupling, we suggest a 540~nm
laser to couple resonantly $2s2p$ and $2s^2(^1S)$ in helium. In such a setup, these two autoionizing states are very weakly ionized
by the laser and can possibly undergo one or more Rabi oscillations within their lifetimes.
By changing the intensity and the wavelength of the laser, there are rich possibilities to manipulate autoionization, and the
effects can be measured in either the absorption or the electron spectra. Unlike long pulses, the large bandwidth of an
attosecond pulse covers the whole autoionization spectrum in a single measurement for each fixed time delay. Such manipulation
would epitomize true attosecond control of electron dynamics. In the meanwhile, the modification of the XUV absorption, if
incorporated with appropriate macroscopic conditions of the medium, would substantially alter the transmitted pulse going through
the medium. This in turn would offer a tool to shape ultrashort light pulses, which serves in the theme of optical coherent
control~\cite{lukin, glover, ignesti10, ignesti11}.

In Sec.~\ref{theory}, our model for the total wavefunction is introduced. It is generalized from the earlier models for long
pulses~\cite{themelis} to short pulses. Then we look at absorption in terms of the frequency-dependent response function, which is
defined as the energy-differential probability density of the absorption through dipole interaction~\cite{gaarde}.
In Sec.~\ref{results}, the calculated spectra are compared with the available experiments~\cite{loh, wang}, and in a test case, the
absorption and the electron spectra are compared and discussed. The conclusions are drawn in Sec.~\ref{conclusion}. Atomic units
($c=\hbar=e=1$) are used throughout Sec.~\ref{theory}, and electron Volts (eV) for energy and attoseconds (as) or femtoseconds (fs)
for time are used in the rest of this paper unless otherwise specified.

\section{Theory}\label{theory}

\subsection{Total wavefunction}\label{theory-total} 

The total wavefunction of a laser-assisted three-level autoionizing system has been formulated in our previous report~\cite{chu11}.
Consider a model atom consisting of the ground state $|g\rangle$ and two autoionizing states $|a\rangle$ and $|b\rangle$, which are
degenerate with the continuum states $|E_1\rangle$ and $|E_2\rangle$ respectively. The atom is exposed to an XUV field that couples
$|g\rangle$ to the first resonance and an IR field that couples the two resonances. The pulses are in arbitrary envelopes and
linearly polarized in the same direction. The total Hamiltonian is $H(t)=H_A+H_I(t)$ where $H_A$ is the atomic Hamiltonian and
$H_I(t)$ is the interaction between the atom and the light. The field is treated as an external disturbance to the system where the
field's own Hamiltonian is excluded. The total wavefunction can be written in the configuration space as
\begin{align}
|\Psi(t)\rangle &= e^{-i E_g t} c_g(t) |g\rangle \notag\\
&+ e^{-i E_X t} \left[ d_a(t) |a\rangle + \int{d_{E_1}(t) |E_1\rangle dE_1} \right] \notag\\
&+ e^{-i E_L t} \left[ d_b(t) |b\rangle + \int{d_{E_2}(t) |E_2\rangle dE_2} \right], \label{total-config}
\end{align}
where $E_g$ is the ground state energy, and $E_X\equiv\omega_X+E_g$ and $E_L\equiv\omega_L+\omega_X+E_g$ are the atomic energies
populated by the pulses. In principle, the energy integrations are over infinite range, but practically only the parts near the
resonances are influenced by the strong coupling, thus worth our concern. Since the fast oscillating parts are factored out in the
exponential terms in Eq.~(\ref{total-config}), the coefficients $c_g(t)$, $d_a(t)$, $d_{E_1}(t)$, $d_b(t)$, and $d_{E_2}(t)$ are
all slowly varying
functions of time. This form is valid when the field is weak to moderate such that the dipole transitions between $|g\rangle$ and
$|a\rangle$ and between $|a\rangle$and $|b\rangle$ are the major dynamics being considered. This means multiphoton effect or
tunneling ionization is to be treated simply as additional depletions of the bound states $|a\rangle$ and $|b\rangle$, where the
rates can be estimated by the theory developed by Ammosov, Delone, and Krainov (ADK theory)~\cite{adk} or the theory developed by
Perelomov, Popov, and Terent'ev (PPT theory)~\cite{ppt}.

Substituting the expansion of $|\Psi(t)\rangle$ into the time-dependent Schr\"{o}dinger equation, first-order coupled equations
for the five coefficients are found, where two coefficients are continuous. As described in~\cite{chu11}, we first apply adiabatic
elimination to $d_{E_1}(t)$ and $d_{E_2}(t)$ to obtain the ``first iteration'' of these two functions of energy and time. They are
then plugged into the other three coupled equations for $c_g(t)$, $d_a(t)$ and $d_b(t)$ (See Eqs.~(12-14) in Ref.~\cite{chu11}),
where these bound-state coefficients can be solved numerically as functions of time. Next, in the ``second iteration'', correcting
the adiabatic elimination, the continuum-state coefficients are solved with the original coupled equations
(See Eqs.~(18) and (19) in Ref.~\cite{chu11}.) Thus, with all the coefficients in Eq.~(\ref{total-config})
calculated, $|\Psi(t)\rangle$ is obtained. In our model, the dipole matrix elements and the Fano parameters are assumed to be
independent of electron energy, and they can be obtained from energy-domain experiments or from separate structure calculations.

The total wavefunction $|\Psi(t)\rangle$ can also be expressed in the eigenstate basis as
\begin{align}
|\Psi(t)\rangle &= e^{-i E_g t} c_g(t) |g\rangle + e^{-i E_X t}  \int{ c_E^{(a)}(t) | \psi_E^{(a)} \rangle dE } \notag\\
&+ e^{-i E_L t}  \int{ c_{E'}^{(b)}(t) | \psi_{E'}^{(b)} \rangle dE' }, \label{total-eig}
\end{align}
where the eigenstates $|\psi_E^{(a)}\rangle$ and $|\psi_{E'}^{(b)}\rangle$ are composed of $|a\rangle$ and $|E_1\rangle$ and of
$|b\rangle$ and $|E_2\rangle$ respectively, as described in Fano's theory~\cite{fano}. When the field is over,
$|\psi_E^{(a)}\rangle$ and $|\psi_{E'}^{(b)}\rangle$ are stationary states of the atomic Hamiltonian, and the coefficients
$c_E^{(a)}(t)$ and $c_E^{(b)}(t)$ are constant of time other than a factor of oscillating phase. The photoelectron spectra of the
two resonances are calculated as $P^{(a)}(E)=|c_E^{(a)}(t_f)|^2$ and $P^{(b)}(E')=|c_{E'}^{(b)}(t_f)|^2$, respectively, where $t_f$
is taken after the pulses are over. In this paper, the time delay $t_0$ is defined between the pulse peaks, and it is positive if
the IR comes after the XUV.

\subsection{Absorption}\label{theory-absorption} 

The frequency-dependent response function $\tilde{S}(\omega)$ for photoabsorption of an atom exposed to a finite-duration external
field has been described recently by Gaarde \textit{et al.}~\cite{gaarde}. Considering an atom in an oscillating electric field,
the response function $\tilde{S}(\omega)$ is defined by
\begin{equation}
\Delta U = \int_0^{\infty}{\omega \tilde{S}(\omega) d\omega} \label{U1}
\end{equation}
where $\Delta U$ is the total energy absorbed by the atom from the field. The response function represents the probability density
for the light of frequency $\omega$ to be absorbed. The value of $\tilde{S}(\omega)$ is positive for absorption and negative for
emission, for a positive $\omega$. In our model, the light-atom interaction term in the Hamiltonian is given by $-D(t)E(t)$ where
$D(t)$ is the electric dipole, and the total absorbed energy through the dipole interaction is
\begin{align}
\Delta U &= \int_{-\infty}^{\infty}{ \frac{dD(t)}{dt} E(t) dt} \label{U2}\\
&= -2\int_{0}^{\infty}{ \omega \: \textnormal{Im} [ \tilde{D}(\omega) \tilde{E}^*(\omega) ] d\omega}, \label{U3}
\end{align}
where $\tilde{D}(\omega)$ and $\tilde{E}(\omega)$ are Fourier transforms of $D(t)$ and $E(t)$ respectively. The convention of
Fourier transform for an arbitrary finite function $f(t)$, from the time domain to the frequency domain, is
\begin{equation}
\tilde{f}(\omega) = \frac{1}{\sqrt{2\pi}} \int_{-\infty}^{\infty}{e^{-i\omega t} f(t) dt}. \label{fourier}
\end{equation}
Note that in Eq.~(\ref{U3}) we have implemented the fact that $D(t)$ and $E(t)$ are all real functions so that 
$\tilde{D}^*(\omega)=\tilde{D}(-\omega)$ and $\tilde{E}^*(\omega)=\tilde{E}(-\omega)$. The final form of the response function, as
extracted from Eq.~(\ref{U1}) and Eq.~(\ref{U3}), is
\begin{equation}
\tilde{S}(\omega) = -2 \: \text{Im} [ \tilde{D}(\omega) \tilde{E}^*(\omega) ] \label{S}.
\end{equation}
As mentioned earlier, $\tilde{S}(\omega)$ is the absorption probability density, which has the same dimension of the photoelectron
profile $P(E)$ defined in Sec.~\ref{theory-total}. The absorption cross section $\tilde{\sigma}(\omega)$ is related to
$\tilde{S}(\omega)$ by
\begin{equation}
\tilde{\sigma}(\omega) = \frac{4\pi \alpha \omega \tilde{S}(\omega)}{|\tilde{E}(\omega)|^2}, \label{xsec}
\end{equation}
where $\alpha$ is the fine structure constant.

With the total wavefunction in Eq.~(\ref{total-eig}), the dipole moment $D(t)$ is given by
\begin{align}
D(t) &= \langle \Psi(t) | \mu | \Psi(t) \rangle \\
&= \left[ e^{i\omega_X t} u_X(t) + e^{i\omega_L t} u_L(t) + c.c. \right] \label{Dt}
\end{align}
where $\mu$ is the electric dipole operator ($\mu=-ez$), and $u_X(t)$ and $u_L(t)$ are
\begin{align}
u_X(t) &\equiv c_g(t) \int{M_E^{(a)} c_E^{(a)*}(t) dE} \\
u_L(t) &\equiv \int{ c_E^{(a)}(t) \int{ M_{E'E}^{(b)} c_{E'}^{(b)*}(t) dE'} dE},
\end{align}
where $M_E^{(a)} \equiv \langle \psi_E^{(a)} |\mu| g \rangle$ and $M_{E'E}^{(b)} \equiv \langle \psi_{E'}^{(b)} |\mu| \psi_E^{(a)}
\rangle$ are the dipole matrix elements between the eigenstates. With Fano's theory~\cite{fano}, $M_E^{(a)}$ and $M_{E'E}^{(b)}$ can
be decomposed to the dipole matrix elements in the configuration space, which are taken as input parameters in our model. Following
Eq.~(\ref{Dt}), the $\tilde{D}(\omega)$ function is
\begin{equation}
\tilde{D}(\omega) = \frac{1}{\sqrt{2\pi}} \int_{-\infty}^{\infty}
{[ e^{i(\omega_X-\omega)t} u_X(t) + e^{i(\omega_L-\omega)t} u_L(t)] dt} \label{Dw}
\end{equation}
where the rotating wave approximation has been applied, provided that $E_b>E_a$. If $E_b<E_a$, $\omega_L$ will be negative, and
$e^{i\omega_L t}u_L(t)$ should be substituted by $e^{-i\omega_L t}u_L^*(t)$ in Eq.~(\ref{Dw}). Since the contributions of the two
light pulses to $\tilde{D}(\omega)$ are quite separate in frequency in the current setup, we can write the response function
involving $u_X(t)$ and $E_X(t)$ as $\tilde{S}_X(\omega)$, and that involving $u_L(t)$ and $E_L(t)$ as $\tilde{S}_L(\omega)$.

In the following calculations, the photoelectron profiles for the two resonances are $P^{(a)}(E)$ and $P^{(b)}(E)$ defined in
Sec.~\ref{theory-total}, and the photoabsorption profiles for XUV and IR are $\tilde{S}_X(\omega)$ and $\tilde{S}_L(\omega)$,
respectively. If the density of the atomic gas is sufficiently low, the profiles given above will be linear to
the actual measured electron and light signals, which can then be assessed at an independent stage.

\section{Results and Discussions}\label{results}

\subsection{Photoabsorption spectra of Ar with single attosecond pulses}\label{case1} 

\begin{figure}[htbp]
\centering
\includegraphics[width=0.5\textwidth]{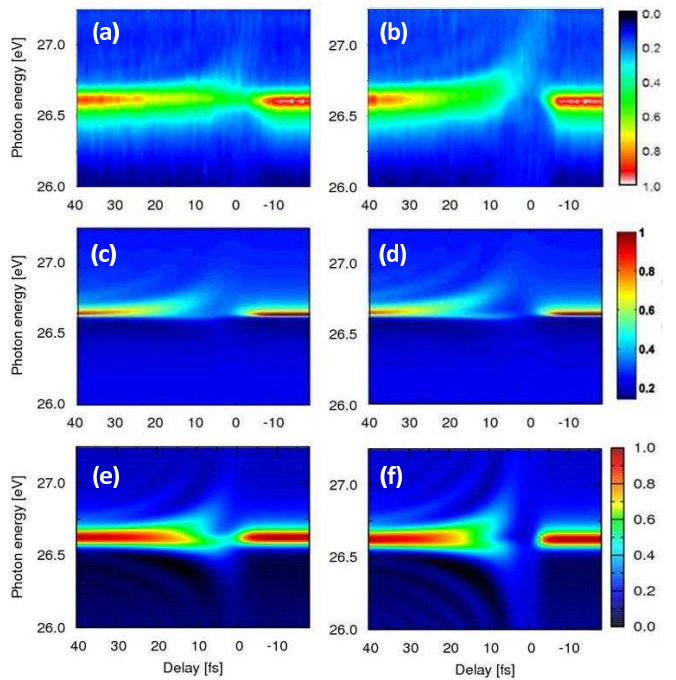}
\caption{(Coler online) Time-delayed transmission spectra of the $3s 3p^6 4d$ resonance in Ar for two IR intensities. The signals
are normalized to 1. The measurements shown in (a-b) are taken from Ref.~\cite{wang}. The simulations in (c-d) are based on the
model in Ref.~\cite{wang} and updated by Zhang~\cite{zhang}.
The present calculations are shown in (e-f). All left panels are for $I_L=0.5$~TW/cm$^2$ and all right panels are for
$I_L=1.0$~TW/cm$^2$.}
\label{fig:case1}
\end{figure}

We now apply the present theory to analyze the experiment by Wang \textit{et al.}~\cite{wang} where an XUV SAP was used to excite
argon atoms to the $3s 3p^6 4p$ autoionizing state in the presence of an intense IR pulse which strongly couples this state with
the $3s 3p^6 4d$ autoionizing state. The weak XUV pulse had duration ($\tau_X$) of 140~as and photon energy ($\omega_X$) covering
20 to 40~eV. The IR pulse had wavelength ($\lambda_L$) of 750~nm ($\omega_L=1.65$~eV) and duration ($\tau_L$) of 7~fs. Two IR peak
intensities ($I_L$), 0.5 and 1.0~TW/cm$^2$ were independently applied. The experiment measured light transmission of the XUV for
different time delays. The experimental results for the two IR intensities
are shown in Fig.~\ref{fig:case1}(a-b). The main conclusion of the experiment is that the transmission peak at the Fano
resonance is depressed when the two pulses overlap, and when the IR lags behind the XUV, the main peak recovers gradually as the
two pulses moved away from each other on the scale of the decay
lifetime of $3s 3p^6 4p$. This result serves to reveal autoionization in the time domain. Note that the delay decreases along
$x$-axis in Fig.~\ref{fig:case1}, so the positive delay (XUV first) are on the left of the plots, which follows the graphing
convention in Ref.~\cite{wang}.

Considering a dilute target gas, the transmission spectrum can be written as
\begin{equation}
\tilde{T}(\omega) = \tilde{T}_0(\omega) \exp [ -\rho \tilde{\sigma}(\omega) L ] \label{transmission}
\end{equation}
where $\tilde{T}(\omega)$ and $\tilde{T}_0(\omega)$ are the light intensity going into and coming out of the medium respectively,
$\rho$ is the gas density, $\tilde{\sigma}(\omega)$ is the absorption cross section for a single atom, and $L$ is the length of the
light path through the medium. The approximation $e^{-x} \approx 1-x$ can be further applied for a low-density gas, where the
final measurement is linearly related to the single-atom response. In this paper, the calculations are made only for single atoms,
where the information of the medium, which is often unavailable, is not needed.

In Ref.~\cite{wang}, the simulation was carried out and recently updated by Zhang~\cite{zhang}. His new results are shown in
Fig.~\ref{fig:case1}(c-d).
Unlike in Ref.~\cite{wang}, the spectrum is now defined as the response
function as in Eq.~(\ref{S}). The dipole moment therein is calculated with the total wavefunction of the earlier model in
Refs.~\cite{bachau, madsen, themelis}, which is equivalent to our model with the exact bound-state coefficients and the first
iteration of the continuum-state coefficients (See Sec.~\ref{theory-total}). As seen in Fig.~\ref{fig:case1}(a-d), the main
discrepancies between the measurements and the updated calculations by Zhang are as follows. First, near $t_0=10$~fs
where the two pulses overlap, the calculation in (d) shows strong peak splitting, but the measurement in (b) has only the
upward-curving branch. The same discrepancy is also seen between (a) and (c). Second, in the measurements, the resonance peak is
depleted much more strongly at the high intensity than at the low intensity, while the simulations in (c-d) do not have this
clear contrast.

The results of the present calculation are shown in Fig.~\ref{fig:case1}(e-f). We include the tunneling ionization rates for
the autoionizing states estimated by the ADK model with the parameters fit to the experiment. For $3s 3p^6 4p$ and
$3s 3p^6 4d$, $C_l=1.6$ and 0.7, and $\alpha=4$ and 5, respectively, where $C_l$ is the ADK coefficient in Ref.~\cite{adk} and
$\alpha$ is the correction parameter in Ref.~\cite{tong}. Our calculations differ from the previous ones, as shown in
Fig.~\ref{fig:case1}(c-d) versus (e-f), in the following ways. First, the resonance peak is heavily smeared by the IR
ionization near the overlap of the pulses. This is especially obvious in Fig.~\ref{fig:case1}(f). Second, the upper branch
of the split peaks tilts upward more than that of the previous simulation, and is closer to what the measurement suggests.
Third, below the resonance energy, there are negative fringes in the spectra curving downward with the delay, which are not seen
in the previous report. By comparing the two models with the experiment, it is obvious that the most important improvement by the
present one is the inclusion of the IR ionization. The one more iteration of the continuum-state coefficients, which is
newly developed in our model, is not essential in this case. Some other detailed features of the result, such as the fringes, are
unexplained so far. On the other hand, they may be washed out in measurements where in principle, signals should be integrated
over a range of intensities.

\subsection{Transient absorption spectra of He with attosecond pulse trains}\label{case2} 

Electromagnetically induced transparency (EIT)~\cite{harris, fleischhauer} has been traditionally studied as a time-independent
phenomenon where a resonance state shows a depleted profile in the absorption spectra when an ac field strongly couples it to
another excited state. While it has been carried out by long pulses and low-lying excited states, a recent work by Loh
\textit{et al.}~\cite{loh} consisting of measurements and simulations investigated the EIT effect initiated by femtosecond
pulses. It employed an 800~nm,
42~fs, 14~TW/cm$^2$ IR pulse to couple the $2s2p(^1P)$ and $2p^2(^1S)$ doubly excited states in helium, and a 30~fs, weak XUV
produced by high-order harmonic generation to probe the IR-coupled system at different time delays. The simulation therein employed
the model developed by Lambropoulos~\cite{lambropoulos} configured for longer pulses. Our analysis for this
experiment illustrates that the present model can be applied to broad-band attosecond pulses as well as longer femtosecond pulses.

\begin{figure}[htbp]
\centering
\includegraphics[width=0.48\textwidth]{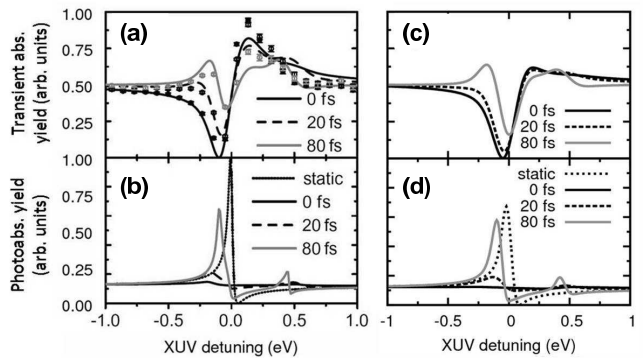}
\caption{The transient absorption spectra (upper panels) and the photoabsorption spectra (lower panels) reported in
Ref.~\cite{loh} (left panels) and calculated by the present model (right panels) for various time delays. The experimental data
reported in Ref.~\cite{loh} is shown in (a) in dots.}
\label{fig:case2}
\end{figure}

Figure~\ref{fig:case2}(a) shows that the theoretical model in Ref.~\cite{loh} can reproduce the major features of the transient
absorption spectra observed in the experiment. Transient absorption spectrum is defined by absorption spectrum subtracted by static
spectrum. Since the IR is rather intense, in Ref.~\cite{loh} it was concluded that the resonance widths of $2s2p$ and $2p^2$ should
be corrected by the IR ionization using the tunneling ionization rates specified in the paper. The transient spectra in
Fig.~\ref{fig:case2}(a) have also been convolved with a 0.18-eV Gaussian function to account for the
spectrometer resolution. A post-pulse which has 33\% of the peak IR intensity, at 88~fs after the main peak is taken into account.
The post-pulse is responsible for the small peak at about 0.45~eV in the absorption spectra in Fig.~\ref{fig:case2}(b) and (d).

In Fig.~\ref{fig:case2}(b), the ``raw'' theoretical photoabsorption spectra from Ref.~\cite{loh} are shown. Here, the ``static''
spectrum means the XUV spectrum measured in the absence of the IR. All the profiles are calculated as the ionization rate
versus XUV detuning in weak-probe limit, given by Eq.~(1) in Ref.~\cite{loh}, which employs the formalism by Madsen
\textit{et al.}~\cite{madsen}. In this formalism, the XUV pulse shape is not considered, and as shown in the spectra, the resonance
peaks are not broadened by the XUV duration. In our model, we calculate the profiles as either the ground state depletion
$1-|c_g(t_f)|^2$ or the total absorption probability $\int{\tilde{S}(\omega)d\omega}$ (which are equivalent theoretically) versus
XUV detuning. Our calculation assumes the XUV peak intensity $I_X=10^{10}$~W/cm$^2$, such that it is in the first-order perturbation
regime. All the XUV and IR electric field parameters are included in our model. The 60-meV bandwidth of the XUV pulse broadens
the resonance peaks in Fig.~\ref{fig:case2}(d).

The theoretical absorption spectra show that when the two pulses overlap, the $2s2p$ resonance peak splits into two peaks moving
outward, corresponding to the Autler-Townes doublet; at the same time, the peaks are highly suppressed due to the IR ionization.
A good agreement has been found between the experiment and both models. The EIT effect by femtosecond pulses, as the goal of this
study, can thus be controlled by overlapping the two pulses with different delays. However, because of the narrow XUV bandwidth,
the absorption is defined for each XUV photon energy, and it is not necessary to carry out the energy distribution in each
measurement by means of the response function $\tilde{S}(\omega)$. In other words, our model is not advantageous over
Lambropoulos' model here. However, for the argon case in Sec.~\ref{case1}, the broadband attosecond XUV retrieves the whole
absorption spectrum in just one measurement, which can be treated in the present model but not in Lambropoulos' model.

\subsection{Electron emission and photoabsorption with single attosecond XUV pulses}\label{case3} 

The outcome of an XUV+IR experiment is usually defined by the time-delayed electron or absorption spectra. In the case of weak XUV
and strong IR, as in EIT, the interaction can be viewed as an XUV probing a strongly IR-coupled atomic system. For the autoionizing
states considered here, the IR can also be viewed as a way to control or manipulate the autoionization dynamics. In the case of
photoabsorption, the single-atom response studied in this work can be implemented in the light field traveling through the medium.
It can then be regarded as a form of optical control or pulse shaping of the XUV by the IR~\cite{glover, ignesti10, ignesti11},
which will be our extended effort of this project in the near future.
Both the electron and the absorption spectra are the observable consequence of the interplay between the electron dynamics and the
optical fields. In this subsection we examine the relation between these two types of measurements.
Before we proceed, note that while the modification of the XUV by the IR has often been studied, the IR is also modified by the XUV
and complimentary ``shaping'' should appear also in the IR spectrum. In the following, by changing the laser wavelength to 540~nm
(it is now actually a visible light instead of an IR) to couple $2s2p(^1P)$ and $2s^2(^1S)$ in helium, we will study the absorption
spectra of both the XUV and the laser, along with the electron spectra. The laser spectra shown therein are not only informative
and interesting, but also assessed as realistically measurable.

In this test case, an XUV pulse with duration $\tau_X=100$~as, peak intensity $I_X=10^{10}$~W/cm$^2$, and central photon energy
$\omega_X=60$~eV is shone on helium to excite it to $2s2p$ at 60.15~eV as well as the ``background''
$1s\epsilon p(^1P)$ continuum states. Because of the wide bandwidth, other nearby doubly excited states, such as $2s3p(^1P)$ and
$2s4p(^1P)$, are also excited as well; they have much narrower widths and are weaker~\cite{domke}, thus disregarded in the present
study. In the meanwhile, a time-delayed laser pulse with wavelength $\lambda_L=540$~nm, duration $\tau_L=9$~fs, and peak intensity 
$I_L=0.7$~TW/cm$^2$ is applied on the system to strongly couples $2s2p$ to the $2s^2(^1S)$ doubly excited state at 57.85~eV
which is below $2s2p$ energetically. The single-ionization binding energy of $2s2p$ and $2s^2$ are 5.3~eV and 7.6~eV, respectively.
With the laser applied, their tunneling ionization rates are negligible, and we expect to see the effect of strong coupling between
them. In the simulation, the dipole matrix elements between $1s^2$ and $2s2p$ and between $2s2p$ and $2s^2$ are 0.038 and
2.17~a.u.~respectively. The Fano parameters are $\Gamma=37$~meV (lifetime is 17~fs) and $q=-2.75$ for $2s2p$, and $\Gamma=0.125$~eV
and $q=1000$ for $2s^2$. The direct ionization path $\langle 1s\epsilon s |\mu| 2s2p \rangle$, which involves transition of two
electrons, is expected to be very weak compared to $\langle 2s^2 |\mu| 2s2p \rangle$, where the ratio is determined by the
$q$-parameter. The calculated electron and absorption spectra are done with various time delays $t_0$ between -15 and 50~fs.

\begin{figure}[htbp]
\centering
\includegraphics[width=0.5\textwidth]{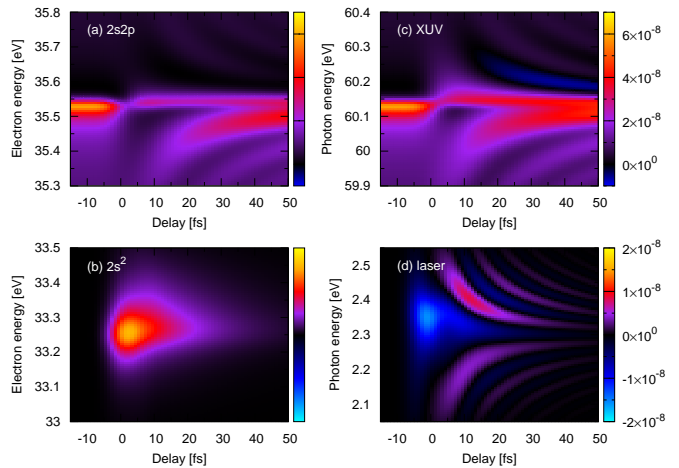}
\caption{(Color online) Photoelectron spectrograms near the (a) $2s2p$ and (b) $2s^2$ resonances, and photoabsorption spectrograms
for the (c) XUV and (d) laser fields, for a fixed laser intensity and wavelength (See text). The negative signals plotted in blue
and cyan colors in (c-d) represent photoemission.}
\label{fig:case3-t0}
\end{figure}

The electron and absorption profiles are shown in Fig.~\ref{fig:case3-t0}. The electron emissions near the two resonances are shown
in Fig.~\ref{fig:case3-t0}(a-b). The photoabsorptions for the XUV and the laser near their central photon energies are shown in
Fig.~\ref{fig:case3-t0}(c-d). Note that the energy scale for all four graphs are the same, which means the signal distributions for
these four calculations can be compared side-by-side. The electron emission of this test case has been reported in details
previously by us~\cite{chu11}; the main task here is to investigate its relation to photoabsorption. Two observations can be drawn
from Fig.~\ref{fig:case3-t0}.

\begin{figure}[htbp]
\centering
\includegraphics[width=0.5\textwidth]{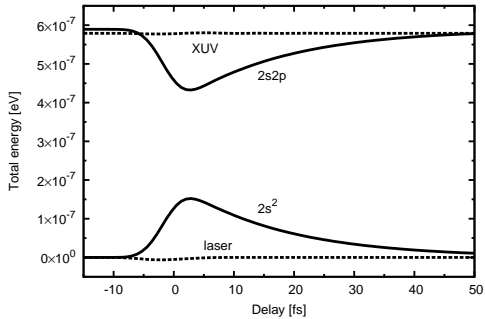}
\caption{The total absorbed and emitted energy across 1-eV spectral range over the resonances versus time delay. The solid curves
are for electron emission and the dashed curves are for photoabsorption.} \label{fig:case3-Etot}
\end{figure}

First, the spectrogram (signal density versus energy and time delay) of the $2s2p$ resonance is almost the same in the electron
and the absorption profile. However, the absorption probability is almost ubiquitously higher because the electrons brought by the
XUV photons from $1s^2$ to $2s2p$ are redistributed between the $2s2p$ and $2s^2$ resonances via the laser coupling. In other
words, all the
electrons shown in Fig.~\ref{fig:case3-t0}(a) and (b), coming out from $2s2p$ and $2s^2$ respectively, have gone through the
absorption shown in Fig.~\ref{fig:case3-t0}(c). This observation is further clarified by energy conservation of the system.
Figure~\ref{fig:case3-Etot} shows the total energies absorbed by the XUV and the laser, as well as the total emitted photoelectron
energies, as functions of time delay. At each time delay, the total energy absorbed by the XUV is calculated by integrating
$\tilde{S}(\omega)$ over the 1-eV range centered at the resonance energy 60.15~eV. Outside this region, the electrons are not
involved in the laser coupling, thus not changing with the time delay. The total emitted electron energy of the $2s2p$ resonance
is calculated by integrating $E |c_E^{(a)}(t_f)|^2$ over the same range. The total energy of laser absorption is small due to its
low photon energy and narrow bandwidth. The total emitted electron energy of the $2s^2$ resonance is obtained by counting energy
over the whole resonance width (Recall that $q=1000$ indicates that the IR coupling does not generate the $1s\epsilon s$
continuum). Figure~\ref{fig:case3-Etot} shows that when the laser pulse appears just after the XUV, a considerable fraction of
energy shifts from $2s2p$ to $2s^2$. As the delay increases, this fraction drops because more electrons decay from $2s2p$ before
the laser couples. The work of laser is to direct electrons to the decay paths of $2s2p$ or $2s^2$, which are separate by only
2.3~eV. Energy conservation is obvious where the total absorbed energy from XUV and from laser equals the total emitted energy
from $2s2p$ and from $2s^2$.

The second observation is the strong correlation between the electron profile of the $2s^2$ resonance and the absorption profile
of the laser for the time delay between 0 and 5~fs, as seen in Figs.~\ref{fig:case3-t0}(b) and (d). The $2s^2$ profile shows a
major peak for which the stimulated emission of laser drives the helium from $2s2p$ to $2s^2$ before $2s2p$
decays. The laser absorption profile has a negative peak of the same strength and in the same time-delay range as the positive peak
of the $2s^2$ electron profile. In other words, the Rabi oscillation during 0 and 5~fs is shown by the electron burst and the laser
emission equivalently. However, as the delay increases, the absorption profile quickly becomes complicated. For each time delay,
the interference fringes are shown in both the absorption and the emission parts of the laser spectra. No obvious explanations of
such structures can be offered at this time.

\subsection{Intensity and wavelength dependence}

The spectrograms shown in Fig.~\ref{fig:case3-t0} contain rich details of how the dressing laser controls the autoionization
dynamics and the absorption of the XUV light. Such details are observable only if electrons and photons are detected with high
spectral resolution because the effects are significant only within the widths of the two autoionizing states. In this subsection,
taking the same test system as in Sec.~\ref{case3}, we examine the electron and the absorption spectra with respect to the
intensity and the detuning of the laser. In the following, Rabi oscillation is quantitatively express by Rabi frequency
\begin{equation}
\Omega (t) = \sqrt{ \left[ D E_0(t) \right]^2 + \delta^2 } \label{rabi_freq}
\end{equation}
where $D$ is the dipole matrix element between the coupled states, $E_0(t)$ is the envelope of the pulse, and $\delta$ is the
detuning of the carrier frequency. For near-resonance condition, pulse area given by
\begin{equation}
A = \int_{-\infty}^{\infty} { \Omega(t) dt}, \label{pulse_area}
\end{equation}
represents the population transfer between the states at the end of the pulse; a thorough transfer from one to the other is
indicated by $A=\pi$. The numerical integration covers the whole temporal range of the coupling pulse.

\begin{figure}[htbp]
\centering
\includegraphics[width=0.5\textwidth]{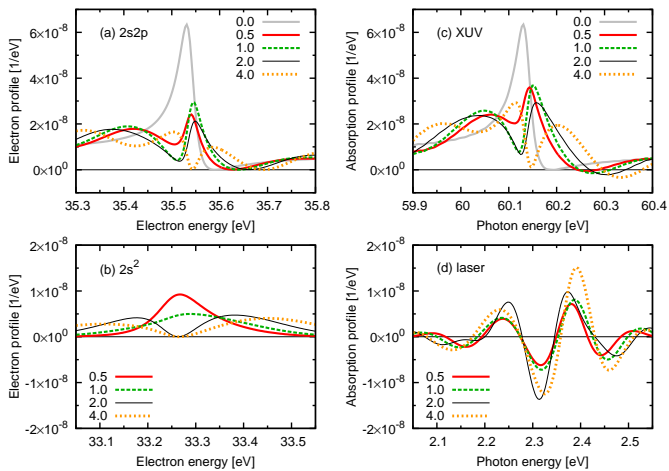}
\caption{(Color online) Photoelectron spectra near the (a) $2s2p$ and (b) $2s^2$ resonances, and photoabsorption spectra for the
(c) XUV and (d) laser fields, for different peak laser intensities (labeled in the units of TW/cm$^2$) and fixed time delay
$t_0=15$~fs. The gray curve is the spectrum measured without the laser. The horizontal black line indicates zero signal.}
\label{fig:case3-IL}
\end{figure}

Figure~\ref{fig:case3-IL} shows the calculated spectra for a fixed time delay $t_0=15$~fs and various laser intensities.
For a typical two-level system, each cycle of Rabi oscillation revives the initial state and changes the phase of its
coefficient by $\pi$. Similarly, for the autoionizing states here, the bound-state coefficients change the phase by $\pi$ after
each cycle. For the lowest laser intensity, i.e., $I_L=0.5$~TW/cm$^2$ and $A=1.34\pi$, the inverse Fano line shape (the
$q$-parameter changes sign) in Fig.~\ref{fig:case3-IL}(a) and the single peak in (b) are formed where a fraction of the wave packet
carrying the changed phase swings back to $2s2p$. This condition has been discussed in details in Ref.~\cite{chu11}. For
$I_L=1$~TW/cm$^2$, or $A=1.89\pi$, the electrons are expressively driven back to $2s2p$. The inverse Fano peak is prominent in
Fig.~\ref{fig:case3-IL}(a) and (c), and the $2s^2$ profile is depleted by about half in Fig.~\ref{fig:case3-IL}(b). As $I_L$
further increases, e.g., $A=2.67\pi$ and $3.78\pi$ for $I_L=2$ and 4~TW/cm$^2$ respectively, in the same laser duration the two
levels undergo more oscillatory cycles before autoionization, which results in more interference fringes within the covered energy
region.

The XUV absorption spectra in Fig.~\ref{fig:case3-IL}(c) are, as explained earlier, almost the same in shape as the $2s2p$ electron
profile but with stronger signals. This implies that the XUV absorption spectra would yield essentially the same information as
that from the electron spectra. Since high-resolution spectra are harder for electrons than for photons, the detailed structures
predicted here are better carried out using photoabsorption spectroscopy. On the other hand, as shown in Fig.~\ref{fig:case3-IL}(c),
at larger $I_L$, the absorption signals near 60.3~eV appear to be negative; it means some amount of energy in that frequency range
is released by the atom in the form of XUV emission. In Fig.~\ref{fig:case3-IL}(d) we further note that the laser absorption
spectra feature an oscillatory pattern with the same nodal points in energy. When the laser is intensified, the nodal points do not
shift, but the amplitude of the pattern increases. This suggests that the absorption and the emission of the laser simultaneously
increase.

\begin{figure}[htbp]
\centering
\includegraphics[width=0.5\textwidth]{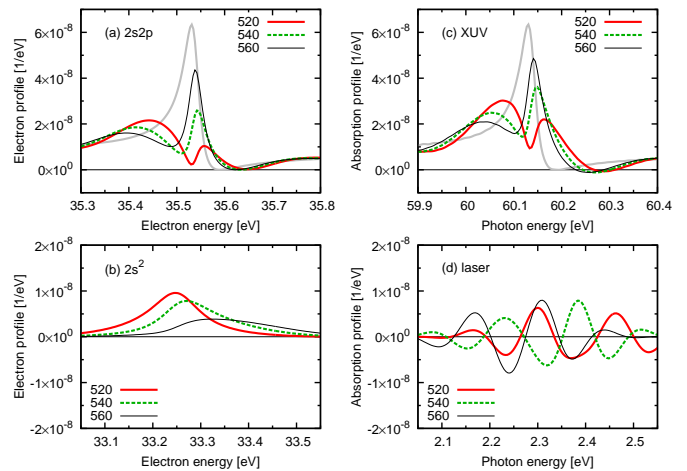}
\caption{(Color online) As Fig.~\ref{fig:case3-IL} but for different laser wavelengths (labeled in nanometers) and fixed peak laser
intensity $I_L=0.7$~TW/cm$^2$ and time delay $t_0=15$~fs.}
\label{fig:case3-det}
\end{figure}

So far we have discussed the special resonance condition where the 540~nm laser resonantly couples $2s2p$ and $2s^2$. If the
wavelength shifts, the detuning should be taken into account, and the spectra are expected to change accordingly.
In Fig.~\ref{fig:case3-det}, two additional wavelengths are plotted, where the detuning is about 0.8~eV above and below the
resonance, respectively, with fixed $I_L=0.7$~TW/cm$^2$. Due to the detuning, the dominantly populated electron energy of the
$2s^2$ resonance is shifted upward or downward as the laser photon energy is lower or higher respectively, thus shifting the $2s^2$
profile as shown in Fig.~\ref{fig:case3-det}(b). The laser absorption patterns for all three wavelengths in
Fig.~\ref{fig:case3-det}(d) are formed by the same
interference mechanism but different carrier frequencies, which consequently show overall shifts in energy but the same geometric
features. The $2s2p$ profile in Fig.~\ref{fig:case3-det}(a) and the XUV profile in (b) are obviously deformed by the detuning.
However, a strong indication of the $2s2p$ resonance energy, at 35.55~eV in electron energy or 60.15~eV in XUV energy, is
unaffected. With the 0.8~eV detuning, its is the wing of the incident laser
profile that couples the two states, where the slope of the wing pickes up additional phase and strength dependence in the Rabi
oscillation, which then further deforms the spectra from the 540~nm case.

\section{Summary and Conclusion}\label{conclusion}

We have developed a theoretical model to calculate the photoabsorption spectra of an atom for an attosecond or femtosecond XUV pulse
in the presence of a moderately intense laser pulse. The XUV pulse pumps the atom to an autoionizing state when a time-delayed laser
pulse couples this state to another autoionizing state. The model is extended from the standard theory of EIT to two autoionizing
states and to ultrashort pulses. The wavefunction in this problem was formulated and applied to calculate photoelectron spectra
previously~\cite{chu11}. In this paper, we generalize the theory to calculate photoabsorption spectra. To illustrate that the
model is applicable to both attosecond and femtosecond pulses, the time-delayed absorption spectra of two available recent
experiments are simulated, one using SAPs on argon~\cite{wang} and the other using APTs on helium~\cite{loh}. For both cases, good
general agreement with experimental data have been found. However, in both experiments, the coupling IR laser severely ionizes the
autoionizing states, and the influence of laser coupling on the resonances is less spectacular. Furthermore, the spectral
resolution in both experiments are still limited and unable to reveal the rich structures of the IR-modified autoionizing profiles.

To highlight the effect of laser coupling, we suggest to use a 540~nm laser to resonantly couple the autoioning $2s2p$ and $2s^2$
states in helium. Because of the large binding energies, ionization by the dressing laser is negligible for both states, and the
coupling effect is maximized. We then investigate the electron profiles of both resonances and the absorption profiles of both the
XUV and the laser pulses. The result shows that the XUV absorption profile near $2s2p$ is almost the same as the corresponding
electron emission from the $2s2p$ resonance, but the absorption signal is higher than the electron signal. The absorption and
electron profiles within the widths of the resonances reveal the resonance features modulated by the dressing pulse.
For such studies, photoabsorption measurements are more favorable since better spectral resolution can be reached with photons.

We should emphasize that the present calculations have been limited to light interactions with single atoms only. For
photoabsorption spectroscopy, light pulses are directed in the gas medium coherently and thus the effect of macroscopic
propagation should be included~\cite{gaarde}. This would offer a practical method to shape the XUV pulse by supplying an
time-delayed intense IR laser in a gas medium, where the propagation effect can be modeled after the high-order harmonic generation
in gases~\cite{jin}.

\begin{acknowledgments}
This work is supported in part by Chemical Sciences, Geosciences and Biosciences Division, Office of Basic Energy Sciences, Office
of Science, U.S. Department of Energy. WCC greatly appreciates the data in Fig.~\ref{fig:case1}(c-d) and the commentary provided by
C.~H.~Zhang.
\end{acknowledgments}


\begin{thebibliography}{}
\bibitem{fano} U.~Fano, Phys. Rev. \textbf{124}, 1866 (1961).
\bibitem{krausz} F.~Krausz and M.~Ivanov, Rev. Mod. Phys. \textbf{81}, 163 (2009).
\bibitem{wickenhauser} M.~Wickenhauser, J.~Burgd\"{o}rfer, F.~Krausz, and M.~Drescher, Phys. Rev. Lett. \textbf{94}, 023002 (2005).
\bibitem{mercouris} Th.~Mercouris, Y.~Komninos, and C.~A.~Nicolaides, Phys. Rev. A \textbf{75}, 013407 (2007).
\bibitem{zhao} Z.~X.~Zhao and C.~D.~Lin, Phys. Rev. A \textbf{71}, 060702 (2005).
\bibitem{chu10} W.-C.~Chu and C.~D.~Lin, Phys. Rev. A \textbf{82}, 053415 (2010).
\bibitem{drescher} M.~Drescher, M.~Hentschel, R.~Kienberger, M.~Uiberacker, V.~Yakovlev, A.~Scrinzi, Th.~Westerwalbeshloh,
U.~Kleineberg, U.~Heinzmann, and F.~Krausz, Nature (London) \textbf{419}, 803 (2002).
\bibitem{gilbertson} S.~Gilbertson \textit{et al.}, Phys. Rev. Lett. \textbf{105}, 263003 (2010).
\bibitem{kitzler} M.~Kitzler, N.~Milosevic, A.~Scrinzi, F.~Krausz, and T.~Brabec, Phys. Rev. Lett. \textbf{88}, 173904 (2002).
\bibitem{hentschel} M.~Hentschel, R.~Kienberger, Ch.~Spielmann, G.~A.~Reider, N.~Milosevic, T.~Brabec, P.~Corkum, U.~Heinzmann,
M.~Drescher, and F.~Krausz, Nature (London) \textbf{414}, 509 (2001).
\bibitem{remetter} T.~Remetter \textit{et al.}, Nature Phys. \textbf{2}, 323 (2006).
\bibitem{goulielmakis} E.~Goulielmakis, V.~S.~Yakovlev, A.~L.~Cavalieri, M.~Uiberacker, V.~Pervak, A.~Apolonski, R.~Kienberger,
U.~Kleineberg, and F.~Krausz, Science \textbf{317}, 769 (2007).
\bibitem{mauritsson} J.~Mauritsson, P.~Johnsson, E.~Mansten, M.~Swoboda, T.~Ruchon, A.~L'Huillier, and K.~J.~Schafer,
Phys. Rev. Lett. \textbf{100}, 073003 (2008).
\bibitem{harris} S.~E.~Harris, J.~E.~Field, and A.~Imamoglu, Phys. Rev. Lett. \textbf{64}, 1107 (1990).
\bibitem{fleischhauer} M.~Fleischhauer, A.~Imamoglu, and J.~P.~Marangos, Rev. Mod. Phys. \textbf{77}, 633 (2005).
\bibitem{arimondo} E.~Arimondo, Prog. Opt. \textbf{35}, 257 (1996).
\bibitem{tamarat} Ph.~Tamarat, B.~Lounis, J.~Bernard, M.~Orrit, S.~Kummer, R.~Kettner, S.~Mais, Th.~Basch\'{e},
Phys. Rev. Lett. \textbf{75} 1514 (1995).
\bibitem{hau} L.~V.~Hau, S.~E.~Harris, Z.~Dutton, and C.~H.~Behroozi, Nature (London) \textbf{397}, 594 (1999).
\bibitem{phillips} D.~F.~Phillips, A.~Fleischhauer, A.~Mair, R.~L.~Walsworth, and M.~D.~Lukin,
Phys. Rev. Lett. \textbf{86}, 783 (2001).
\bibitem{vudyasetu} P.~K.~Vudyasetu, R.~M.~Camacho, and J.~C.~Howell, Phys. Rev. Lett. \textbf{100}, 123903 (2008).
\bibitem{loh} Z.-H.~Loh, C.~H.~Greene, and S.~R.~Leone, Chem. Phys. \textbf{350}, 7 (2008).
\bibitem{lambropoulos} P.~Lambropoulos and P.~Zoller, Phys. Rev. A \textbf{24}, 379 (1981).
\bibitem{bachau} H.~Bachau, P.~Lambropoulos, and R.~Shakeshaft, Phy. Rev. A \textbf{34}, 4785 (1986).
\bibitem{madsen} L.~B.~Madsen, P.~Schlagheck, and P.~Lambropoulos, Phys. Rev. Lett. \textbf{85}, 42 (2000).
\bibitem{themelis} S.~I.~Themelis, P.~Lambropoulos, and M.~Meyer, J. Phys. B: At. Mol. Opt. Phys. \textbf{37}, 4281 (2004).
\bibitem{chu11} W.-C.~Chu, S.-F.~Zhao, and C.~D.~Lin, Phys. Rev. A \textbf{84} 033426 (2011).
\bibitem{wang} H.~Wang \textit{et al}, Phys. Rev. Lett. \textbf{105}, 143002 (2010).
\bibitem{gaarde} M.~B.~Gaarde, C.~Buth, J.~L.~Tate, and K.~J.~Schafer, Phys. Rev. A \textbf{83}, 013419 (2011).
\bibitem{lukin} M.~D.~Lukin, Rev. Mod. Phys. \textbf{75}, 457 (2003).
\bibitem{glover} T.~E.~Glover \textit{et al.}, Nature Phys. \textbf{6}, 69 (2010).
\bibitem{ignesti10} E.~Ignesti, R.~Buffa, L.~Fini, E.~Sali, M.~V.~Tognetti, and S.~Cavalieri, Phys. Rev. A \textbf{81}, 023405 (2010).
\bibitem{ignesti11} E.~Ignesti, R.~Buffa, L.~Fini, E.~Sali, M.~V.~Tognetti, and S.~Cavalieri, Phys. Rev. A \textbf{83}, 053411 (2011).
\bibitem{adk} M.~V.~Ammosov, N.~B.~Delone, and V.~P.~Krainov, Sov. Phys. JETP \textbf{64}, 1191 (1986).
\bibitem{ppt} A.~M.~Perelomov, V.~S.~Popov, M.~V.~Terent'ev, Sov. Phys. JETP \textbf{23}, 924 (1966); \textbf{24}, 207 (1967).
\bibitem{zhang} C.~H.~Zhang (private communication).
\bibitem{tong} X.~M.~Tong and C.~D.~Lin, J. Phys. B: At. Mol. Opt. Phys. \textbf{38}, 2593 (2005).
\bibitem{domke} M.~Domke, K.~Schulz, G.~Remmers, G.~Kaindl, and D.~Wintgen, Phys. Rev. A \textbf{53}, 1424 (1996).
\bibitem{jin} C.~Jin, A.-T.~Le, and C.~D.~Lin, Phys. Rev. A \textbf{83}, 023411 (2011).
\end{thebibliography}
\end{document}